\documentclass[preprint,showpacs,preprintnumbers,amsmath,amssymb,prb]{revtex4}

\usepackage{graphicx}
\usepackage{dcolumn}
\usepackage{bm}
\usepackage{longtable}

\begin{document}

\title{Inhomogeneous spectral moment sum rules for the retarded Green function and self-energy of strongly correlated electrons or ultracold fermionic atoms in optical lattices}

\author{J.~K.~Freericks}
\homepage{http://www.physics.georgetown.edu/~jkf}
 \affiliation{Department of Physics, Georgetown University,
Washington, D.C. 20057}

\author{V.~Turkowski} 
\affiliation{Department of Physics and Nanoscience and Technology Center, 
   University of Central Florida, Orlando, Florida 32816}

\date{\today}

\begin{abstract}
Spectral moment sum rules are presented for the inhomogeneous many-body problem described by the fermionic Falicov-Kimball
or Hubbard models. These local sum rules allow for arbitrary hoppings, site energies, and interactions.  They can be employed to quantify the accuracy of numerical solutions to the inhomogeneous many-body problem like strongly correlated multilayered devices, ultracold atoms in an optical lattice with a trap potential, strongly correlated systems that are disordered, or systems with nontrivial spatial ordering like a charge density wave or a spin density wave. We also show how the spectral moment sum rules determine the asymptotic behavior of the Green function, self-energy, and dynamical mean field, when applied to the dynamical mean-field theory solution of the many body problem.  In particular, we illustrate in detail how one can dramatically reduce the number of Matsubara frequencies needed to solve the Falicov-Kimball model, while still retaining high precision, and we sketch how one can incorporate these results into Hirsch-Fye quantum Monte Carlo solvers for the Hubbard (or more complicated) models. Since the solution of inhomogeneous problems is significantly more time consuming than periodic systems, efficient use of these sum rules can provide a dramatic speed up in the computational time required to solve the many-body problem. We also discuss how these sum rules behave in nonequilibrium situations as well, where the Hamiltonian has explicit time dependence due to a driving field or due to the time-dependent change of a parameter like the interaction strength or the origin of the trap potential.
\end{abstract}



\maketitle

\section{Introduction}

Spectral moments are integrals of powers of frequency multiplied by the corresponding spectral function and integrated over all frequency.  They can reveal important information about the structure and spread of the spectral function and, in some cases, can also reveal interesting information about different many-body correlation functions. In theory, knowledge of all spectral moments allows one to reconstruct the function, but that procedure is well-known to be unstable and is not commonly used in numerical calculations. The spectral moments also correspond to derivatives of the Green functions with respect to relative time (either real time or imaginary time), evaluated at the point where the relative time is zero.  As such, the spectral moments provide information about the relative-time dependence, when expanded as a Taylor series in time.

Spectral moment sum rules for the many-body problem were investigated in 1967 by Harris and Lange~\cite{harris_lange} shortly after the Hubbard model\cite{hubbard} was introduced. In that work, one can find the moment sum rules for the first three moments of the retarded Green function of the Hubbard model and various approximations like the alloy analogy problem (which is equivalent to an inhomogeneous Falicov-Kimball model\cite{falicov_kimball}). They also developed a strong-coupling projection method to find spectral moments within each of the different Hubbard bands.  This approach is an approximate one, as the projection operator is developed in a power series of the hopping divided by the interaction strength. Shortly thereafter, Nolting\cite{nolting} applied the spectral moment sum rules to develop approximations for the momentum-dependent Green function that have two poles, with the weights and locations of the poles fixed by the corresponding sum rules.  This approach has been developed quite extensively, and applied to a variety of different problems including dynamical mean-field theory\cite{geipel_nolting,nolting_borgiel,potthoff,potthoff2,eskes}. That work examined ferromagnetic and antiferromagnetic long-range order in the Hubbard model, different approximation schemes for perturbation theory that produce the correct strong-coupling limit, and extended the sum rules to the retarded self-energy.  The approach has also been applied to photoemission and inverse photoemission\cite{kornilovitch}, where it was recognized that the moments of the so-called lesser and greater Green functions play an important role.  Finally, Harris and Lange's projection technique was applied to the lesser and greater Green functions to examine spectral properties of the Hubbard model\cite{randeria}.

Most of that work has had as its focus using the sum rules to develop different approximations, or to learn semiquantitative features of the many-body problem.  But there is another application of spectral moment sum rules that is quite important for computational work.  The sum rules allow, in certain circumstances, for numerically exact computations to be benchmarked against the sum rules.  Steven White applied this to the Hubbard model in two dimensions\cite{white}, and Deisz, Hess, and Serene applied it to the fluctuation-exchange approximation\cite{deisz_hess_serene} at the Matsubara frequencies. The sum rules have been generalized to nonequilibrium situations for the Green functions\cite{turkowski_freericks1} and self-energies\cite{turkowski_freericks2}.  In the case when one applies a spatially uniform, but time-dependent, electric field to the system, it turns out that many of the low-order moments are time-independent, even though the Hamiltonian has explicit time-dependence due to the field being turned on at a specific time (or because the field has nontrivial time dependence). Nonequilibrium dynamical mean-field theory\cite{freericks_turkowski_zlatic,turkowski_freericks_book,freericks} can solve the many-body problem exactly, and the sum rules are employed to benchmark the quality of the solutions\cite{dod_ugc_2006}.

Finally, spectral moment sum rules, when viewed as a Taylor series expansion in time, are now being employed to improve both the speed and the quality of Hirsch-Fye quantum Monte Carlo approaches\cite{hirsch_fye} for solving the impurity problem in dynamical mean-field theory\cite{jarrell}. In particular, recent work\cite{blumer0,blumer1,blumer2} has shown that employing the exact Taylor series expansions for short imaginary times allows one to use much larger Trotter error, yet achieve high accuracy with the Hirsch-Fye algorithm, so that it is competitive with continuous-time-based approaches\cite{continuous_time}.

In this work, we want to extend the spectral-moment sum rules for the retarded Green function and retarded self-energy to inhomogeneous cases, which are becoming more important, and where one can find similar improvement in the efficiency of impurity solvers for use in inhomogeneous dynamical mean-field theory problems.  There are four main thrusts of work in the inhomogeneous many-body problem currently: (i) examining the properties of strongly correlated multilayers because of their potential for quantum-mechanical engineering of device properties\cite{freericks_book}; (ii) examining the properties of ultracold atoms on optical lattices but spatially confined by a magnetic or optical trap\cite{optical_lattice}; (iii) examining the properties of a strongly correlated material that is disordered; and (iv) examining properties of strongly correlated materials with inhomogeneous spatial ordering (like a charge or spin density wave). In the multilayer problem, most solutions have relied on approximate techniques for the Hubbard model, or have examined simplified models like the Falicov-Kimball model (recently, however, the numerical renormalization group has been applied to multilayer Hubbard systems\cite{rosch_hubbard_multilayer,freericks_hubbard_multilayer}).  In the cold atom problem, little work has been applied to Green function-based techniques (although this is increasing); here the spectral-moment sum rules could aid both in benchmarking and in improving the accuracy and efficiency of numerical algorithms.  The strongly correlated material with disorder problem has had many different techniques applied to it, but most of them are approximate in one way or another, so understanding the quality of the approximations is important. Less work has taken place in ordered phase calculations, especially for sum rules when the system is spatially ordered; these results can be immediately examined with the results presented here. In all cases, use of sum rules can help provide quantitative data to analyze how accurate the numerical solutions are. Finally, inhomogeneous systems are likely to be studied within the nonequilibrium context. Here, we also generalize the nonequilibrium sum rules to the inhomogeneous environment and we consider a wide range of different nonequilibrium contexts.

The remainder of this contribution is organized as follows: in Section II, we discuss the formalism for deriving the spectral moments in equilibrium; in Section III, we generalize to nonequilibrium cases; in Section IV, we discuss different applications of sum rules within dynamical mean-field theory; and in Section V, we present the conclusions and summary.

\section{\label{sec:level2} Formalism for deriving spectral moments in equilibrium}
\label{Formalism}

The equilibrium Hamiltonian for the inhomogeneous Falicov-Kimball\cite{falicov_kimball} and  Hubbard\cite{hubbard} models
 can be written in the following unified form:
\begin{eqnarray}
{\mathcal H}=-\sum_{ij}t_{ij}c_{i}^{\dagger}c^{}_{j}
-\sum_{ij}t_{ij}^{f}f_{i}^{\dagger}f^{}_{j}
-\sum_{i}\mu_{i}c_{i}^{\dagger}c^{}_{i}
-\sum_{i}\mu_{fi}f_{i}^{\dagger}f^{}_{i}
+\sum_{i}U_{i}f_{i}^{\dagger}f^{}_{i}c_{i}^{\dagger}c^{}_{i}. \label{eq: ham}
\end{eqnarray}
Here $c_i^\dagger$ ($c^{}_i$) are the creation (annihilation) operators for a conduction electron at site $i$ and $f^\dagger_i$ and $f^{}_i$ are the corresponding operators for a localized electron at site $i$; in the case of the Hubbard model, the $c$-electrons are the spin-up electrons and the $f$-electrons are the spin-down electrons. The hopping matrix is denoted by $-t_{ij}^{}$ and $-t_{ij}^f$ for the $c$- and $f$-electrons, respectively; for the Falicov-Kimball model we have $t^f=0$, while for the Hubbard model we have $t^f=t$ (one can also consider an asymmetric Hubbard model with $0<t^f\ne t$).  Note that the hopping matrix need not be translationally invariant, the only requirement is that it is Hermitian. The local chemical potentials are denoted by $\mu_i=\mu-V_i$ and $\mu_{fi}=\mu_f-V_{fi}$ for the conduction and localized electrons with $V_i$ and $V_{fi}$ the corresponding local potentials ($\mu_i=\mu_{fi}$ for the Hubbard model in a vanishing magnetic field; $\mu_{fi}=0$ for the Falicov-Kimball model, since it does not enter into the conduction electron moments), and the local interaction between different particles is $U_i$.  In the case of disorder, one often averages the Hamiltonian, and different measurable operator expectation values, over the given disorder distribution for the different parameters that are disordered. We will not discuss the details of how to treat those kinds of problems here. The sum rules we derive would be for one quenched disorder configuration (corresponding to a particular choice of parameters in the Hamiltonian), and could subsequently be averaged with respect to the given disorder distribution, if desired. Note that this Hamiltonian is time independent, and we have no net current flow, so it corresponds to an equilibrium problem (in other words, it is in the ``slow limit'' if the potentials correspond to an electric field, where the charge rearranges itself into a static redistribution in response to the potential rather than allowing current to flow).  We will also examine a wide class of nonequilibrium cases below.

We do not make any assumptions about the translational invariance of any of the parameters that enter the Hamiltonian, so the Green function will generically depend on two spatial coordinates rather than on their difference. But in equilibrium, the system does have time-translation invariance, so we can describe the Green functions with a single frequency by making a temporal Fourier transform.
The retarded Green function is defined to be
\begin{equation}
G_{ij}^{R}(t_{1},t_{2}) =-i\theta (t_{1}-t_{2}){\rm Tr} e^{-\beta\mathcal{H}} \left\{
c_{i}(t_{1}),c_{j}^{\dagger}(t_{2})\right\}_+/\mathcal{Z}, \label{eq: retarded_green_def}
\end{equation}
in the time representation, with $\mathcal{Z}={\rm Tr} \exp(-\beta\mathcal{H})$ the partition function, $\beta$ the inverse temperature, and $\{A,B\}_+=AB+BA$ the anticommutator. The creation and annihilation operators are in the (equilibrium) Heisenberg representation, where $\mathcal{O}(t)=\exp(i\mathcal{H}t)\mathcal{O}\exp(-i\mathcal{H}t)$ for any operator $\mathcal{O}$; in this representation, the time-translation invariance is easy to show due to the invariance properties of the trace and the fact that the Hamiltonian commutes with itself. The frequency representation for the retarded Green function is
\begin{equation}
 G_{ij}^{R}(\omega)=\int dt e^{i\omega t}G_{ij}^R(t,0).
\label{eq: green_retarded_w}
\end{equation}
The spectral function is then defined to be
\begin{eqnarray}
A_{ij}^{R}(\omega) =-\frac{1}{\pi} {\rm Im}G_{ij}^{R}(\omega), \label{eq: spectral_retarded_def}
\end{eqnarray}
and the spectral moments become
\begin{eqnarray}
\mu_{n}^{R}({\bf R}_{i},{\bf
R}_{j})&=&\int_{-\infty}^{\infty}d\omega \omega^{n}
A_{ij}^{R}(\omega). \label{eq: spectral_moment_def}
\end{eqnarray}
Similar to the homogeneous case\cite{turkowski_freericks1,turkowski_freericks2}, one can easily show that
\begin{equation}
\mu_{n}^{R}({\bf R}_{i},{\bf R}_{j})= -2{\rm Im} \left[
i^{n}\frac{\partial^{n}}{\partial t^{n}} G_{ij}^{R}(t,0)
\right]_{t=0^{+}};  \label{eq: spectral_moment_time}
\end{equation}
note that this representation uses the fact that $\theta(0)=1/2$, so the factor of 2 is needed in order to make $2\theta(0)=1$ and have the correct normalization.
Using the Heisenberg equation of motion, one can relate the time derivatives to commutators with the Hamiltonian, to obtain
\begin{eqnarray}
\mu_{n}^{R}({\bf R}_{i},{\bf R}_{j})&=&{\rm Re} \langle
\{L^{n}c_{i}(0),c_{j}^{\dagger}(0)\}_+ \rangle  , \label{eq: spectral_moment_commutator}
\end{eqnarray}
where $L^{n}\mathcal{O}=[...[[\mathcal{O},\mathcal{H}],\mathcal{H}]...\mathcal{H}]$ is the multiple commutation
operator, performed $n$ times. In Eq.~(\ref{eq: spectral_moment_commutator}), we used a shorthand notation, where the angular brackets denote the trace over all states weighted by the density matrix $\exp(-\beta\mathcal{H})/\mathcal{Z}$.

Evaluating the commutators is straightforward, but tedious.  The results are
\begin{eqnarray}
\mu_{0}^{R}({\bf R}_{i},{\bf R}_{j},T)&=&\delta_{ij}, \label{eq: mu0}\\
\mu_{1}^{R}({\bf R}_{i},{\bf R}_{j},T)&=&-t_{ij}-\delta_{ij}(\mu_{i}-U_{i}n_{fi}),\label{eq: mu1}\\
\mu_{2}^{R}({\bf R}_{i},{\bf
R}_{j},T)&=&\sum_{l}t_{il}t_{lj}+(\mu_{i}-U_in_{fi})t_{ij}+t_{ij}(\mu_{j}-U_jn_{fj})\nonumber\\
&+&\delta_{ij}[(\mu_{i}-U_in_{fi})^2+U_i^2n_{fi}(1-n_{fi})],\label{eq: mu2}
\end{eqnarray}
\begin{eqnarray}
\mu_{3}^{R}({\bf R}_{i},{\bf
R}_{j},T)&=&-\sum_{kl}t_{ik}t_{kl}t_{lj}\nonumber\\
&-&(\mu_{i}-U_{i}n_{fi})\sum_{l}t_{il}t_{lj}
-\sum_{l}t_{il}(\mu_{l}-U_{l}n_{fl})t_{lj}
-\sum_{l}t_{il}t_{lj}(\mu_{j}-U_{j}n_{fj})
\nonumber \\
&-&(\mu_i-U_in_{fi})^2t_{ij}-(\mu_i-U_in_{fi})t_{ij}(\mu_j-U_jn_{fj})-t_{ij}(\mu_j-U_jn_{fj})^2\nonumber\\
&-&U_i^2n_{fi}(1-n_{fi})t_{ij}-U_it_{ij}U_j\left [\langle f^\dagger_if^{}_if^\dagger_jf^{}_j\rangle-n_{fi}n_{fj}\right ] -t_{ij}U_j^2n_{fj}(1-n_{fj})\nonumber\\
&-&\delta_{ij}\left [ (\mu_i-U_in_{fi})^3+3U_i^2\mu_in_{fi}(1-n_{fi})-U_i^3n_{fi}(1-n_{fi})(1+n_{fi})\right ]
\nonumber \\
&+&\delta_{ij}U_{i}\sum_{l,m} \left[ t_{mi}^{f}t_{lm}^{f}\langle
f_{l}^{\dagger}f_{i}\rangle +t_{im}^{f}t_{ml}^{f}\langle
f_{i}^{\dagger}f_{l}\rangle -2t_{im}^{f}t_{li}^{f}\langle
f_{l}^{\dagger}f_{m}\rangle \right]
\nonumber \\
&+&\delta_{ij}U_{i}\sum_{l}\left( \mu_{l}^{f}-\mu_{i}^{f} \right) 
\left[ t_{li}^{f}\langle f_{l}^{\dagger}f_{i}\rangle +t_{il}^{f}\langle f_{i}^{\dagger}f_{l}\rangle \right]
+\delta_{ij}U_{i}^{2}\sum_{l}t_{il}^{f}\langle f_{i}^{\dagger}f_{l}\rangle
\nonumber \\
&-&\delta_{ij}U_{i}\sum_{l}U_{l}\left[ t_{li}^{f}\langle f_{l}^{\dagger}f_{i}c_{l}^{\dagger}c_{l} \rangle 
                                          +t_{il}^{f}\langle f_{i}^{\dagger}f_{l}c_{l}^{\dagger}c_{l} \rangle \right]
+U_{i}U_{j}\left[ t_{ji}^{f}\langle f_{j}^{\dagger}f_{i}c_{j}^{\dagger}c_{i} \rangle 
                                          + t_{ij}^{f}\langle f_{i}^{\dagger}f_{j}c_{j}^{\dagger}c_{i} \rangle \right] ,
\nonumber \\
\label{eq: mu3}
\end{eqnarray}
where $n_{fi}=\langle f_{i}^{\dagger}f^{}_{i}\rangle$; this result corrects a factor of 2 error in the third line of Eq.~(29) of Ref.~\onlinecite{turkowski_freericks2}.   Note, that in cases where we have inhomogeneity arising from the spatial long-range order (say charge or spin density wave order, for example), then the Hamiltonian is actually translationally invariant, but the sum rules have inhomogeneous results due to the explicit evaluation of the different expectation values (the charge density will vary from site to site in a charge density wave, for example).  This then gives rise to spatially inhomogeneous moments.

Next, we examine the retarded self-energy moments.
To begin, we start with
the Dyson equation in the real space for the frequency-dependent Green function and self-energy:
\begin{equation}
G_{ij}^{R}(\omega )= G_{ij}^{R0}(\omega )+\sum_{kl}G_{ik}^{R0}(\omega
)\Sigma_{kl}^{R}(\omega )G_{lj}^{R}(\omega ),
\label{eq: dyson_equilib}
\end{equation}
where $G^{R0}_{ij}(\omega)$ is the noninteracting retarded Green function on the lattice. When the frequency $\omega$ is large enough, all Green functions and self-energies become purely real (on the infinite-dimensional hypercubic lattice, there can be an exponentially small imaginary part since the bandwidth is infinite, but this plays little role for large frequencies), and by using the spectral formulas,
\begin{eqnarray}
G_{ij}^{R}(\omega )=-\frac{1}{\pi}\int_{-\infty}^{\infty}d\omega^\prime \frac{{\rm
Im}G_{ij}^{R}(\omega^\prime)}{\omega -\omega^\prime+i\delta}, \label{eq: G_spectral}
\end{eqnarray}
\begin{eqnarray}
\Sigma_{ij}^{R}(\omega )=-\frac{1}{\pi}\int_{-\infty}^{\infty}d\omega^\prime
\frac{{\rm Im}\Sigma_{ij}^{R}(\omega^\prime)}{\omega -\omega^\prime+i\delta}
+\Sigma_{ij}^R(\omega=\infty), \label{eq: sigma_spectral}
\end{eqnarray}
we can expand the functional dependence of the Green function and self-energies in terms of the corresponding moments
(since the denominators in the integrals never vanish when the numerators are nonzero) yielding
\begin{eqnarray}
G_{ij}^{R}(\omega )&=&\sum_{m=0}^{\infty}\frac{ \mu_{m}^{R}({\bf
R}_{i},{\bf R}_{j})}{\omega^{m+1}}, \label{eq: G_expansion}
\\
\Sigma_{ij}^{R}(\omega )&=&\Sigma_{ij}^{R}(\omega =\infty
)+\sum_{m=0}^{\infty}\frac{C_{m}^{R}({\bf R}_{i},{\bf
R}_{j})}{\omega^{m+1}}, \label{eq: sigma_expansion}
\end{eqnarray}
where
\begin{eqnarray}
C_{m}^{R}({\bf R}_{i},{\bf R}_{j})=-\frac{1}{\pi}\int_{-\infty}^{\infty}d\omega
\omega^{m} {\rm Im}\Sigma_{ij}^{R}(\omega), \label{eq: sigma_moment_def}
\end{eqnarray}
are the spectral moments for the self-energy [$\Sigma_{ij}^{R}(\omega =\infty)$ is a real constant equal to the large-frequency limit of the self-energy].
The expansions in Eqs.~(\ref{eq: G_expansion}) and (\ref{eq: sigma_expansion}) (and a similar expansion for the noninteracting retarded Green function) are substituted into the Dyson equation in Eq.~(\ref{eq: dyson_equilib}) and then one equates powers of $1/\omega$ to find
\begin{eqnarray}
\mu_{0}^{R}({\bf R}_{i},{\bf R}_{j})&=&{\tilde \mu}_{0}^{R}({\bf R}_{i},{\bf R}_{j})
\label{eq: relation0}\\
\mu_{1}^{R}({\bf R}_{i},{\bf R}_{j})&=&{\tilde \mu}_{1}^{R}({\bf
R}_{i},{\bf R}_{j})+{\tilde \mu}_{0}^{R}({\bf R}_{i},{\bf
R}_{l})\Sigma_{lm}^{R}(\omega =\infty )\mu_{0}^{R}({\bf
R}_{m},{\bf R}_{j}) ,
\label{eq: relation1}\\
\mu_{2}^{R}({\bf R}_{i},{\bf R}_{j})&=&{\tilde \mu}_{2}^{R}({\bf
R}_{i},{\bf R}_{j})+{\tilde \mu}_{0}^{R}({\bf R}_{i},{\bf
R}_{l})\Sigma_{lm}^{R}(\omega =\infty )\mu_{1}^{R}({\bf
R}_{m},{\bf R}_{j}) \nonumber \\
&~&+{\tilde \mu}_{0}^{R}({\bf R}_{i},{\bf R}_{l})C_{0}^{R}({\bf
R}_{l},{\bf R}_{m})\mu_{0}^{R}({\bf R}_{m},{\bf R}_{j})
\nonumber \\
&~&+{\tilde \mu}_{1}^{R}({\bf R}_{i},{\bf
R}_{l})\Sigma_{lm}^{R}(\omega =\infty )
\mu_{0}^{R}({\bf R}_{m},{\bf R}_{j}),\label{eq: relation2}\\
\mu_{3}^{R}({\bf R}_{i},{\bf R}_{j})&=&{\tilde \mu}_{3}^{R}({\bf
R}_{i},{\bf R}_{j})+{\tilde \mu}_{0}^{R}({\bf R}_{i},{\bf
R}_{l})\Sigma_{lm}^{R}(\omega =\infty )\mu_{2}^{R}({\bf
R}_{m},{\bf R}_{j}) \nonumber \\
&~&+{\tilde \mu}_{0}^{R}({\bf R}_{i},{\bf R}_{l})C_{0}^{R}({\bf
R}_{l},{\bf R}_{m})\mu_{1}^{R}({\bf R}_{m},{\bf R}_{j})
\nonumber \\
&~& +{\tilde \mu}_{0}^{R}({\bf R}_{i},{\bf R}_{l})C_{1}^{R}({\bf
R}_{l},{\bf R}_{m}) \mu_{0}^{R}({\bf R}_{m},{\bf R}_{j})
\nonumber \\
&~& +{\tilde \mu}_{1}^{R}({\bf R}_{i},{\bf
R}_{l})\Sigma_{lm}^{R}(\omega =\infty )\mu_{1}^{R}({\bf
R}_{m},{\bf R}_{j})
\nonumber \\
&~& +{\tilde \mu}_{1}^{R}({\bf R}_{i},{\bf R}_{l})C_{0}^{R}({\bf
R}_{l},{\bf R}_{m}) \mu_{0}^{R}({\bf R}_{m},{\bf R}_{j})
\nonumber \\
&~&+{\tilde \mu}_{2}^{R}({\bf R}_{i},{\bf
R}_{l})\Sigma_{lm}^{R}(\omega =\infty )\mu_{0}^{R}({\bf
R}_{m},{\bf R}_{j}), \label{eq: relation3}
\end{eqnarray}
where the matrix ${\tilde \mu}_{n}^{R}({\bf R}_{i},{\bf R}_{j})$
is the $n$th spectral moment of the noninteracting retarded Green function on the
lattice. Those noninteracting moments are found from Eqs.~(\ref{eq: mu0})--(\ref{eq: mu3}) with $U_i=0$.
Substituting the explicit values of the moments into Eqs.~(\ref{eq: relation0})--(\ref{eq: relation3}) finally yields the self-energy moments:
\begin{equation}
\Sigma_{ij}^{R}(\omega =\infty )=\delta_{ij}U_{i}n_{fi} ,
\label{eq: sigma_infty}
\end{equation}
\begin{equation}
C_{0}^{R}({\bf R}_{i},{\bf R}_{j})=
\delta_{ij}U_{i}^2n_{fi}(1-n_{fi}). \label{eq: c0}
\end{equation}
Note that the algebra required to arrive at these results is nontrivial.  In cases where $i$ and $j$ are farther apart than the range of the hopping matrix (which is often chosen to be nonzero only for nearest neighbors) many moments are identically zero, but nontrivial cancellations are required to ensure that all of the off-diagonal moments vanish (no local approximation has been made for the self-energy here---these results hold in all dimensions).
The first self-energy moment is:
\begin{eqnarray}
C_{1}^{R}({\bf R}_{i},{\bf R}_{j})&=&
\delta_{ij}U_{i}^{2}n_{fi}(1-n_{fi})[U_i(1-n_{fi})-\mu_{i}]
\nonumber \\
&~&+\delta_{ij}U_{i}\sum_{l,m} \left[ t_{mi}^{f}t_{lm}^{f}\langle
f_{l}^{\dagger}f_{i}\rangle +t_{im}^{f}t_{ml}^{f}\langle
f_{i}^{\dagger}f_{l}\rangle -2t_{im}^{f}t_{li}^{f}\langle
f_{l}^{\dagger}f_{m}\rangle \right]
\nonumber \\
&~&+\delta_{ij}U_{i}\sum_{l}\left( \mu_{l}^{f}-\mu_{i}^{f} \right) 
\left[ t_{li}^{f}\langle f_{l}^{\dagger}f_{i}\rangle +t_{il}^{f}\langle f_{i}^{\dagger}f_{l}\rangle \right]
+\delta_{ij}U_{i}^{2}\sum_{l}t_{il}^{f}\langle f_{i}^{\dagger}f_{l}\rangle
\nonumber \\
&~&-\delta_{ij}U_{i}\sum_{l}U_{l}\left[ t_{li}^{f}\langle f_{l}^{\dagger}f_{i}c_{l}^{\dagger}c_{l} \rangle 
                                          +t_{il}^{f}\langle f_{i}^{\dagger}f_{l}c_{l}^{\dagger}c_{l} \rangle \right]
-U_i t_{ij}U_j[\langle f^\dagger_if^{}_if^\dagger_jf^{}_j\rangle -n_{fi}n_{fj}]
\nonumber \\
&~&+U_{i}U_{j}\left[ t_{ji}^{f}\langle f_{j}^{\dagger}f_{i}c_{j}^{\dagger}c_{i} \rangle 
                                          + t_{ij}^{f}\langle f_{i}^{\dagger}f_{j}c_{j}^{\dagger}c_{i} \rangle \right] ,
 \label{eq: c1}
\end{eqnarray}
which contains an off-diagonal term when one is in finite dimensions (it is local in infinite dimensions due to the scaling of the hopping matrix element with dimension).

The expression for the
nonhomogeneous Green function moments Eqs.~(\ref{eq:
mu0})-(\ref{eq: mu3}) and for the self-energy moments Eqs.~(\ref{eq:
sigma_infty})-(\ref{eq: c1}) are rather general. In particular they
can be used to evaluate the moments in the case of a particular
(quenched) configuration of disorder. Then we would average over some disorder distribution [say, $P(V_{i})$ for diagonal disorder].  This procedure requires one to perform calculations over a range of different disorder distributions and then perform the averaging.  If the system tends to self-average, not too many specific configurations would be needed, but if there is interesting physics arising from rare regions of the distributions, many calculations would be needed, and these calculations can get to be rather lengthy.

\section{Generalization to nonequilibrium situations}

One rather general form for the nonequilibrium Hubbard-Falicov-Kimball Hamiltonian is
\begin{eqnarray}
{\mathcal H}(t)&=&-\sum_{ij}t_{ij}(t)c_{i}^{\dagger}c^{}_{j}
-\sum_{ij}t_{ij}^{f}(t)f_{i}^{\dagger}f^{}_{j}
-\sum_{i}\mu_{i}(t)c_{i}^{\dagger}c^{}_{i}
-\sum_{i}\mu_{fi}(t)f_{i}^{\dagger}f^{}_{i}\nonumber\\
&+&\sum_{i}U_{i}(t)f_{i}^{\dagger}f^{}_{i}c_{i}^{\dagger}c^{}_{i}. \label{eq: ham_noneq}
\end{eqnarray}
In this case, we have added time dependence to all of the parameters in the Hamiltonian,
but have not introduced any additional forms of interaction within the Hamiltonian.  Nevertheless,
this generalization allows for a rather rich class of nonequilibrium problems to be studied.  For example, if we are examining a multilayered device with electronic charge reconstruction~\cite{millis} (where the potentials $V_i$ and $V_i^f$ are determined by an additional semiclassical Poisson equation), and we use a vector potential to describe an external electric field that drives current through the system~\cite{freericks_book,freericks_multilayer_thermal}, then we would have a time dependent hopping determined by the Peierls substitution~\cite{peierls}. If we want to examine an interaction quench, as is often studied in cold atom systems, we would have a (typically harmonic) trapping potential $V_i$ and $V_i^f$ and the interaction $U_i$ would become time dependent switching from one value for early times to another value for later times~\cite{eckstein} such as would occur near a Feshbach resonance if the bias magnetic field is changed from one value to another (for some experiments, the potentials or hopping could also change when the coupling changes); the switching could be sudden as in a rapid quench, or adiabatic, with a slowly varying change, or anything in between. In addition, within the cold-atom picture, we could imagine creating time-dependent trap potentials $V_i(t)$ and $V_i^f(t)$.  This would allow us to examine what would happen if we applied an impulse to the atomic cloud, or if we shifted the origin of the harmonic potential from one spatial location to another, and then examined how the center-of-mass oscillates and damps back to the thermal state, or to some nonthermal steady state.  Finally, we could examine the so-called Bragg spectroscopy experiment, where the optical lattice potential amplitude is oscillated with some set frequency and one observes things like the change in the momentum distribution after the system is driven for a certain period of time, or the change in the double occupancy.  In this case, the hopping, the local potentials, and the interaction could all become time dependent, but can still be described by the general form of our Hamiltonian.

There are a few subtle points to keep in mind with these nonequilibrium
problems.  For example, in the case of a multilayered system with
electronic charge reconstruction, the charge reconstruction is
created in the distant past, so it corresponds to an equilibrium
static potential which does not contribute to any flow of current.
The field that drives the current is described by a time-dependent
vector potential, which is turned on at a particular time.  One can
examine the transient current flow or the steady-state current flow
by examining how the system responds to the external field.

Given this general form for the Hamiltonian, we next need to derive
the sum rules. We work in a Heisenberg picture, because the operator
algebra for the time-dependent creation and annihilation operators,
at equal times, is unchanged from the standard fermionic commutation
relations. The only difference from the nonequilibrium derivations
worked out previously\cite{turkowski_freericks1,turkowski_freericks2}
is that here we need to work in real space for all calculations,
because there is no translational invariance.  In order to evaluate the
expression for the nonequilibrium spectral moments, it is convenient
to introduce the relative $t=t_{1}-t_{2}$ and the average
$T=(t_{1}+t_{2})/2$ time coordinates for the retarded Green function
Eq.~(\ref{eq: retarded_green_def}). In this case, the physical time
at which one would likes to calculate the moments will correspond to the average time
$T$, and the Fourier transform to frequency space must be
performed with respect to the relative time $t$. Then, one can define the
nonequilibrium Green function moments by generalizing the expression in
Eq.~(\ref{eq: spectral_moment_def}):
\begin{eqnarray}
\mu_{n}^{R}({\bf R}_{i},{\bf
R}_{j},T)&=&-\frac{1}{\pi}\int_{-\infty}^{\infty}d\omega \omega^{n} {\rm Im}
G_{ij}^{R}(T,\omega). \label{eq: spectral_moment_defT}
\end{eqnarray}
In a similar way, one can define the nonequilibrium moments for the
self-energy:
\begin{eqnarray}
C_{m}^{R}({\bf R}_{i},{\bf
R}_{j},T)=-\frac{1}{\pi}\int_{-\infty}^{\infty}d\omega \omega^{m}
{\rm Im}\Sigma_{ij}^{R}(T,\omega) \label{eq: sigma_moment_defT}
\end{eqnarray}
(for more details, see Ref.~\onlinecite{turkowski_freericks2}). By
using these equations, one can easily show that the expressions for the
nonequilibrium retarded Green function moments in Eqs.~(\ref{eq:
mu0})--(\ref{eq: mu3}) remain unchanged in the nonequilibrium
case, except the model parameters and the operator expectation values are replaced by their
time-dependent forms: $U_{i}\rightarrow U_{i}(T)$, $\mu_{i}\rightarrow \mu_{i}(T)$ and
$n_{fi}\rightarrow n_{fi}(T)$. Also, the operators in the correlation functions
in Eq.~(\ref{eq: mu3}) are in the Heisenberg representation at the average time $T$, at which
the spectral moment is calculated. Similarly, one can show that the
expressions for the large frequency self-energy in Eq.~(\ref{eq:
sigma_infty}) and for the zeroth self-energy moment in Eq.~(\ref{eq:
c0}) remain the same in the nonequilibrium case. However, the
expression for the first moment in Eq.~(\ref{eq: c1}) will acquire an
additional term proportional to the second derivative of the large
frequency limit of the self-energy $\Sigma_{ij}(T,\omega =\infty)$.
Namely, in the nonequilibrium case one finds:
\begin{eqnarray}
C_{1}^{R}({\bf R}_{i},{\bf R}_{j},T)&=&
\delta_{ij}U_{i}^{2}(T)n_{fi}(T)[1-n_{fi}(T)][U_i(T)\{1-n_{fi}(T)\}-\mu_{i}(T)]
\nonumber \\
&~&+\delta_{ij}U_{i}(T)\sum_{l,m} \left[ t_{mi}^{f}(T)t_{lm}^{f}(T)\langle
f_{l}^{\dagger}(T)f_{i}(T)\rangle +t_{im}^{f}(T)t_{ml}^{f}(T)\langle
f_{i}^{\dagger}(T)f_{l}(T)\rangle \right .\nonumber\\
&~&\left .
~~~~~~~~~~~~~~~~~~~~~
-2t_{im}^{f}(T)t_{li}^{f}(T)\langle
f_{l}^{\dagger}(T)f_{m}(T)\rangle \right]
\nonumber \\
&~&+\delta_{ij}U_{i}(T)\sum_{l}\left( \mu_{l}^{f}(T) -\mu_{i}^{f}(T)\right) 
\left[ t_{li}^{f}(T)\langle
f_{l}^{\dagger}(T)f_{i}(T)\rangle +t_{il}^{f}(T)\langle f_{i}^{\dagger}(T)f_{l}(T)\rangle \right]\nonumber\\
\nonumber \\
&~&+\delta_{ij}U_{i}^{2}(T)\sum_{l}t_{il}^{f}(T)\langle
f_{i}^{\dagger}(T)f_{l}(T)\rangle
+\delta_{ij}\frac{1}{4}\frac{\partial^{2}[U_i(T)n_{fi}(T) ]}{\partial T^{2}}\nonumber\\
&~&-\delta_{ij}U_{i}(T)\sum_{l}U_{l}(T)\left[ t_{li}^{f}(T)\langle f_{l}^{\dagger}(T)f_{i}(T)c_{l}^{\dagger}(T)c_{l}(T) \rangle
\right.
\nonumber \\
&~&\left.
~~~~~~~~~~~~~~~~~~~~~~~~~~~~~
+t_{il}^{f}(T)\langle f_{i}^{\dagger}(T)f_{l}(T)c_{l}^{\dagger}(T)c_{l}(T) \rangle \right]
\nonumber \\
&-&U_i(T) t_{ij}(T)U_j(T)[\langle f^\dagger_i(T)f^{}_i(T)f^\dagger_j(T)f^{}_j(T)\rangle -n_{fi}(T)n_{fj}(T)]
\nonumber \\
&~&+U_{i}(T)U_{j}(T)\left[ t_{ji}^{f}(T)\langle f_{j}^{\dagger}(T)f_{i}(T)c_{j}^{\dagger}(T)c_{i}(T) \rangle 
\right.
\nonumber \\
&~&\left.
~~~~~~~~~~~~~~~~~~~~
+ t_{ij}^{f}(T)\langle f_{i}^{\dagger}(T)f_{j}(T)c_{j}^{\dagger}(T)c_{i}(T) \rangle \right] 
. \label{eq: c1nonequilibrium}
\end{eqnarray}
For most nonequilibrium problems we would consider, except for an adiabatically changing interaction, or an amplitude oscillation of the optical lattice, the derivative term would vanish almost everywhere.

\section{Application of spectral moment sum rules to dynamical mean-field theory}

There are two immediate applications of spectral moment sum rules within dynamical mean-field theory.  The first is to use them to evaluate the high-frequency asymptotic behavior exactly and then supplement the high frequency results by numerical calculations at low frequencies, and the second, closely related, is to use them to evaluate the short imaginary time behavior exactly and supplement with long-time numerical calculations. The latter has been already discussed within the context of the Hirsch-Fye quantum Monte Carlo algorithm for solving the impurity problem in dynamical mean-field theory and the results there show great promise as a means to improve the accuracy and the efficiency of calculations\cite{blumer1,blumer2}. The basic idea (at half-filling) is that the curvature, which grows sharply with increasing $U$, immediately determines the short-time behavior for the Green function. Because of the sharp dependence on $\tau$, one would need to use a very small discretization step for the QMC to accurately describe such behavior, which would then be very costly in terms of computational time.  Instead, one uses a coarser grid, but before performing the Fourier transform to Matsubara frequencies, one simply creates a finer grid and uses the short time relations to find the behavior close to $\tau=0$, and uses simple interpolation for other time values (a shape preserving spline would work well in this context).  Then the Fourier transformation will much more accurately reflect the true behavior of the system, and all of the high-frequency structure will be properly recovered, so that the QMC can be used to determine the low-frequency data where it is most accurate. (It has already been demonstrated that this approach is competitive with other QMC techniques such as the continuous-time algorithm.) Issues of accuracy and efficiency will become increasingly important for inhomogeneous dynamical mean-field theory problems (like multilayers or ultracold atoms in a trap) because one needs to solve an impurity problem at each inequivalent lattice site of the inhomogeneous system. This can range from tens to hundreds of impurity solvers for multilayered systems to many thousands or more for ultracold atomic systems in a trap. We won't discuss the application within quantum Monte Carlo approaches further here, and instead will concentrate on examining a different application, which is to solve the dynamical mean-field theory for the Falicov-Kimball model with fixed local chemical potentials on the lattice sites.  This approach works equally well (with appropriate modifications) for the two-site approximation to the Hubbard model\cite{potthoff_2site} in the insulating phase.

The dynamical mean-field theory for the Falicov-Kimball model is well established in the literature\cite{brandt_mielsch,freericks_review}, so we just present the relevant formulas.
Starting from a local self-energy $\Sigma_i(i\omega_n)$, one must solve the Dyson equation for the local Green function [$G_{ii}(i\omega_n)$]
\begin{equation}
 \sum_k\left \{ \left [ i\omega_n+\mu_i-\Sigma_i(i\omega_n)\right ]\delta_{ik}+t_{ik}\right \}G_{kj}(i\omega_n)=\delta_{ij}.
\label{eq: dmft_dyson}
\end{equation}
Here $\omega_n=\pi  (2n+1)/\beta$ is the fermionic Matsubara frequency.
When the system is homogeneous, a Fourier transformation allows this problem to be solved immediately.  When one has inhomogeneity in one dimension, as in multilayered structures, the local Green function (of an infinite device) can be found from the so-called quantum zipper algorithm\cite{potthoff_nolting,freericks_book} which expresses the Green function in terms of continued fractions.  For finite cold atom systems, one can use LAPACK (or sparse matrix) routines to perform numerical matrix inversions to find the local Green function\cite{atkinson,costi}. Whatever the technique, we assume that one can solve the Dyson equation to determine the local Green function.  Next, the effective medium $\mathcal{G}^0_i$ and dynamical mean field $\lambda_i$ are extracted via the scalar equations
\begin{equation}
 \mathcal{G}^0_i(i\omega_n)=\left \{ [G_{ii}(i\omega_n)]^{-1}+\Sigma_i(i\omega_n)\right \}^{-1}.
\label{eq: dmft_g0}
\end{equation}
and
\begin{equation}
 \lambda_i(i\omega_n)=i\omega_n+\mu_i-[\mathcal{G}^0_i(i\omega_n)]^{-1}.
\label{eq: lambda}
\end{equation}
[All quantities in Eqs.~(\ref{eq: dmft_g0}) and (\ref{eq: lambda}) are scalar quantities---there are no matrix operations here.]
Then one calculates the filling of the local electrons $n_{fi}=\mathcal{Z}_{1i}/(\mathcal{Z}_{0i}+\mathcal{Z}_{1i})$ with
\begin{equation}
 \mathcal{Z}_{0i}=e^{\beta\mu_i/2}\prod_{n=-\infty}^{\infty}\frac{i\omega_n+\mu_i-\lambda_i(i\omega_n)}{i\omega_n}
\label{eq: z0}
\end{equation}
and
\begin{equation}
 \mathcal{Z}_{1i}=e^{\beta(\mu_i-U_i)/2}e^{\beta\mu_{fi}}\prod_{n=-\infty}^{\infty}\frac{i\omega_n+\mu_i-\lambda_i(i\omega_n)-U_i}{i\omega_n}.
\label{eq: z1}
\end{equation}
Now we find the new local Green function
\begin{equation}
 G_{ii}(i\omega_n)=\frac{1-n_{fi}}{i\omega_n+\mu_i-\lambda_i(i\omega_n)}+\frac{n_{fi}}{i\omega_n+\mu_i-\lambda_i(i\omega_n)-U_i},
\label{eq: dmft_imp_solve}
\end{equation}
and finally extract the local self-energy
\begin{equation}
 \Sigma_i(i\omega_n)=[\mathcal{G}^0_i(i\omega_n)]^{-1}-G_{ii}^{-1}(i\omega_n).
\label{eq: dmft_dyson2}
\end{equation}
These equations are then iterated until they converge.
The conduction electron filling satisfies
\begin{equation}
 \rho_{ci}=\frac{1}{\beta}\sum_{n=-\infty}^\infty G_{ii}(i\omega_n);
\label{eq: cond_el_fill}
\end{equation}
this result is not part of the dynamical mean-field theory iteration, but it is needed if one wants to update the chemical potential during the iterations to achieve a particular electron filling. Note that this summation is ill-defined and needs to be properly regularized (see below).

Typically, one chooses a set number of Matsubara frequencies, usually with an energy cutoff many multiples of the noninteracting electron bandwidth, and solves the (now finite) set of equations for the Green functions and self-energies by iteration starting from $\Sigma_i=0$.  One can try to include the effects of the neglected tails of the summations and infinite products to improve the accuracy and minimize the effect of the energy cutoff.  By employing the exact sum rules, one can make this procedure work well.  We describe this process next.

First recall that we have already proven that the Matsubara frequency Green function and self-energy satisfy
\begin{equation}
G_{ii}(i\omega_n)=\sum_{m=0}^\infty\frac{\mu^R_m({\bf R}_i,{\bf R}_i)}{(i\omega_n)^{m+1}},
\label{eq: gf_moment_mats}
\end{equation}
and
\begin{equation}
 \Sigma_i(i\omega_n)=\Sigma_i(\infty)+\sum_{m=0}^\infty\frac{C^R_m({\bf R}_i,{\bf R}_i)}{(i\omega_n)^{m+1}},
\label{eq: sigma_moment_mats}
\end{equation}
when the Matsubara frequency is large $|\omega_n|\gg |U_{\rm max}|+W_{int}$, where $W_{int}$ is the half bandwidth (in real frequency) of the interacting density of states (valid only for finite-dimensional systems).  These results follow from the definition of the spectral moments, and the spectral formula for the Green function and self-energy. Using these two relations, substituting into Eq.~(\ref{eq: dmft_g0}), and recalling the definition in Eq.~(\ref{eq: lambda}), produces
\begin{equation}
 \lambda_i(i\omega_n)=\frac{1}{2}\frac{1}{i\omega_n}+\frac{1}{2}\frac{U_in_{fi}-\mu_i}{(i\omega_n)^2}+\mathcal{O}\left ( \frac{1}{(i\omega_n)^3}\right ).
\label{eq: lambda_exp}
\end{equation}
This asymptotic expansion along with the expansion in Eq.~(\ref{eq: gf_moment_mats}) will allow us to treat the tails in the summation for the conduction-electron filling in Eq.~(\ref{eq: cond_el_fill}) and in the infinite products needed for the localized-electron filling in Eqs.~(\ref{eq: z0}) and (\ref{eq: z1}).

We imagine taking an energy cutoff $E_c$ which is larger than the interacting density of states half bandwidth.  Using that cutoff to determine the explicit Matsubara frequencies solved in the numerical implementation of dynamical mean-field theory, we examine only Matsubara frequencies with $|\omega_n|<E_c$.  This defines a cutoff integer $n_c$ corresponding to the Matsubara frequency closest to $E_c$ but lying below it. The tail for the conduction electron filling $\sum_{-\infty}^{-n_c-2}G_{ii}(i\omega_n)/\beta +\sum_{n_c+1}^\infty G_{ii}(i\omega_n)/\beta$ can now be evaluated analytically for the first four moments that we have calculated.  Define the approximate Green function via
\begin{equation}
 G^{approx}_{ii}(i\omega_n)=\frac{1}{i\omega_n}+\frac{\mu^R_1({\bf R}_i,{\bf R}_i)}{(i\omega_n)^2}
+\frac{\mu^R_2({\bf R}_i,{\bf R}_i)}{(i\omega_n)^3}+\frac{\mu^R_3({\bf R}_i,{\bf R}_i)}{(i\omega_n)^4}.
\label{eq: g_approx}
\end{equation}
Then, the identity
\begin{equation}
 f(-\mu)=\frac{1}{1+\exp(-\beta\mu)}=\frac{1}{\beta}\sum_{n=-\infty}^{\infty}\frac{1}{i\omega_n+\mu}
\label{eq: fd}
\end{equation}
allows us to evaluate infinite sums of inverse powers of $i\omega_n$ noting that the sum of $1/i\omega_n$ requires special regularization which is given by the result in Eq.~(\ref{eq: fd}). By taking derivatives, and evaluating at $\mu=0$, we immediately learn that
\begin{eqnarray}
\frac{1}{\beta} \sum_n\frac{1}{i\omega_n}&=&f(0)=\frac{1}{2},\\
\frac{1}{\beta}\sum_n\frac{1}{(i\omega_n)^2}&=&-f^\prime(0)=-\frac{\beta}{4},\\
\frac{1}{\beta}\sum_n\frac{1}{(i\omega_n)^3}&=&\frac{1}{2}f^{\prime\prime}(0)=0,\\
\frac{1}{\beta}\sum_n\frac{1}{(i\omega_n)^4}&=&-\frac{1}{6}f^{\prime\prime\prime}(0)=\frac{\beta^3}{48}.
\end{eqnarray}
Substituting into the expression for the conduction-electron filling then yields
\begin{equation}
 \rho_c=\frac{1}{2}-\frac{\beta}{4}\mu^R_1({\bf R}_i,{\bf R}_i)+\frac{\beta^3}{48}\mu^R_3({\bf R}_i,{\bf R}_i)+\frac{1}{\beta}\sum_{n=-n_c-1}^{n_c}[G_{ii}(i\omega_n)-G^{approx}_{ii}(i\omega_n)].
\end{equation}
The summation, which needs to be evaluated numerically, vanishes rapidly for large $\omega_n$, so the filling can be computed quite accurately.

Now we show how to evaluate the infinite products by properly taking into account the asymptotic limits using the spectral sum rules.
Substituting the results from Eq.~(\ref{eq: lambda_exp}) into Eqs.~(\ref{eq: z0}) and (\ref{eq: z1}) allow for the local electron filling to be computed. First note that
\begin{equation}
\mathcal{Z}_{0i}=e^{\beta\mu_i/2}\prod_{n=n_c+1}^\infty\left | 1+\frac{\mu_i}{i\omega_n}-\frac{1}{2}\frac{1}{(i\omega_n)^2}-\frac{1}{2}\frac{U_in_{fi}-\mu_i}{(i\omega_n)^3}\right |^2
\prod_{n=0}^{n_c}\left |
\frac{1}{\mathcal{G}^0_i(i\omega_n)i\omega_n}\right |^2,
\label{eq: inf_prod_extrap0}
\end{equation}
and
\begin{equation}
\mathcal{Z}_{1i}=e^{\beta(\mu_i-U_i)/2}e^{\beta\mu_{fi}}\prod_{n=n_c+1}^\infty\left | 1+\frac{\mu_i-U_i}{i\omega_n}-\frac{1}{2}\frac{1}{(i\omega_n)^2}-\frac{1}{2}\frac{U_in_{fi}-\mu_i}{(i\omega_n)^3}\right |^2
\prod_{n=0}^{n_c}\left | \left [\frac{1}{\mathcal{G}^0_i(i\omega_n)}-U_i\right ]\frac{1}{i\omega_n}\right |^2.
\label{eq: inf_prod_extrap1}
\end{equation}
The infinite products that have an infinite number of terms are approximated by rewriting the infinite product as the exponential of the sum of the logarithm of the individual terms.  Replacing the sum by an integral (valid when the temperature is much smaller than the interacting half bandwidth) and converting the integral over frequency to an integral over $z=1/\omega$ yields
\begin{equation}
 \prod_{n=n_c+1}^\infty\left | 1+\frac{a}{i\omega_n}+\frac{b}{(i\omega_n)^2}+\frac{c}{(i\omega_n)^3}\right |^2\approx
\exp \left [ \frac{\beta}{2\pi }\int_0^{1/E_c}\frac{dz}{z^2}\ln\left ( \{1+bz^2\}^2+z^2\{a-cz^2\}^2\right )\right ],
\label{eq: inf_prod_integral}
\end{equation}
where we use $a=\mu_i$ for $\mathcal{Z}_{0i}$, $a=\mu_i-U_i$ for $\mathcal{Z}_{1i}$, $b=-1/2$ for both, and $c=-(U_in_{fi}-\mu_i)/2$ for both. This then allows for an accurate evaluation of the filling using just the small set of numerical data generated for $|n|<n_c$.

\begin{figure}[t]
\centerline{\includegraphics [width=3.2in, angle=0, clip=on]  {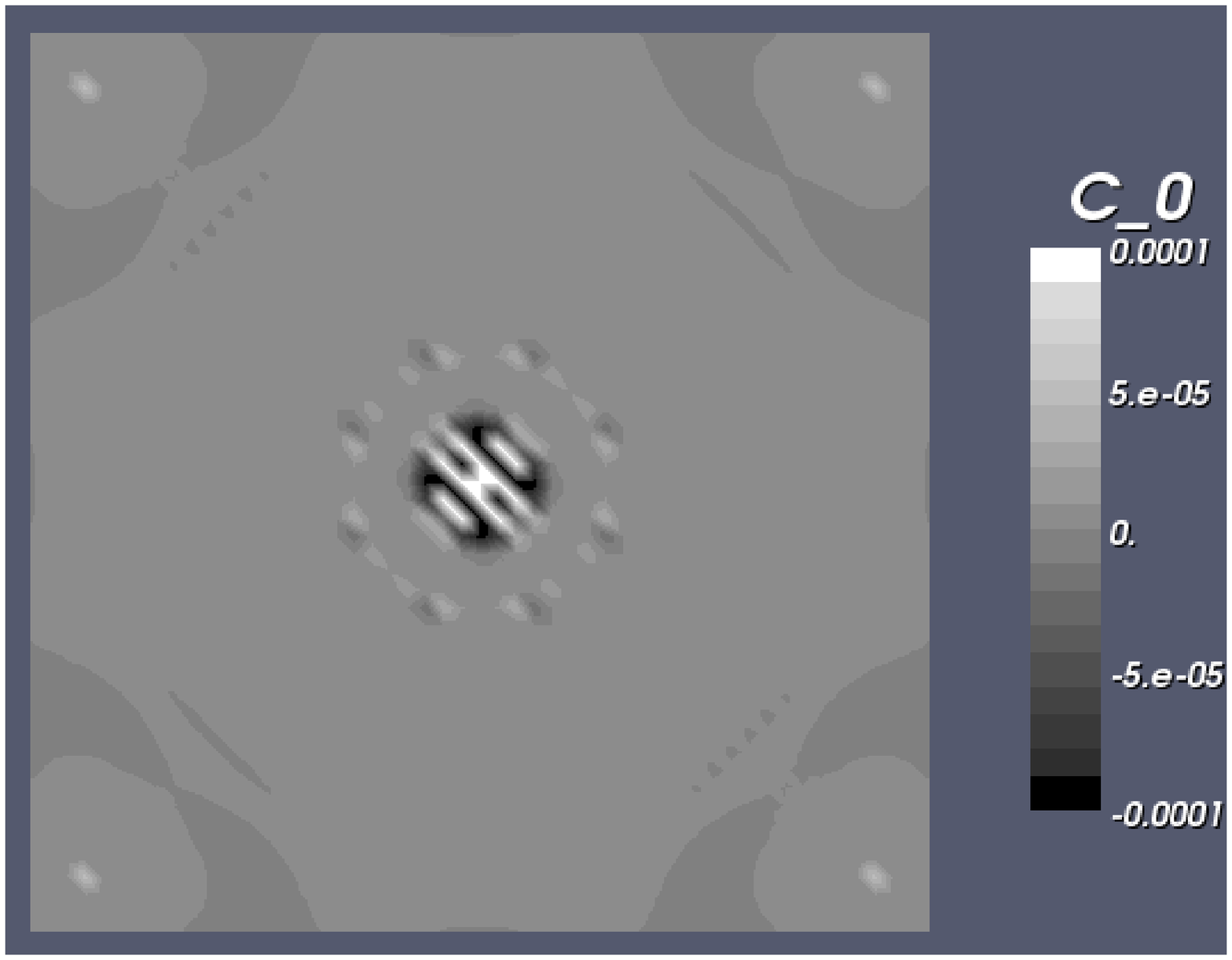}
\includegraphics [width=3.2in, angle=0, clip=on]  {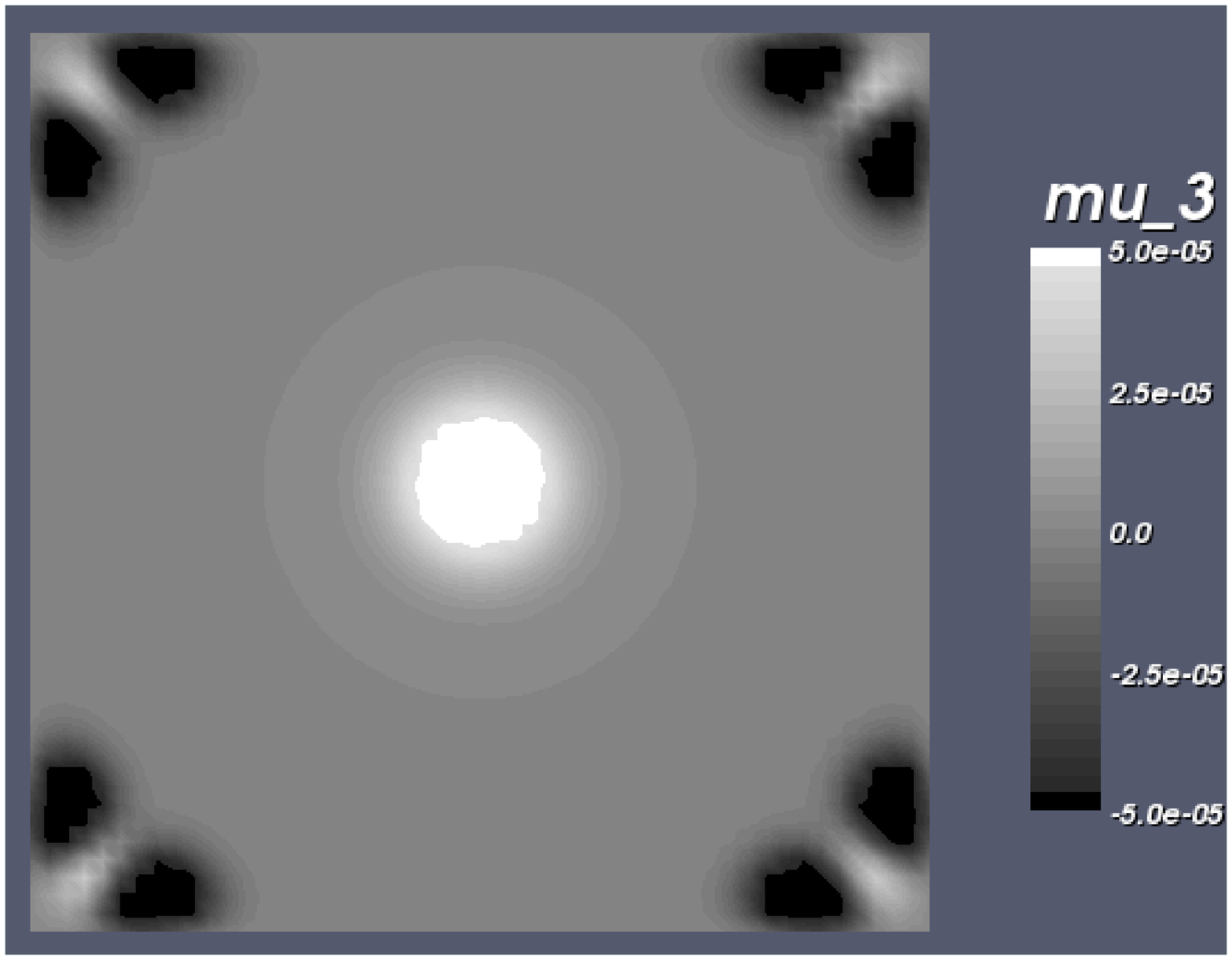}}
\caption[]{
False grayscale image of the relative error of the (left) zeroth self-energy
moment and the (right) third spectral function moment for a trapped atoms system confined in harmonic traps with
a length scale of 30 lattice units, $U=5$, $1/\beta=0.2$ and $51\times 51$ lattice sites. Note that the scale for the figures is less than the full range of deviations to pick up the fine structure far from the center of the lattice.  At the center of the lattice, the error is maximal and equal to approximately 1.5\% for the zeroth
self-energy moment and approximately 0.01\% for the spectral function.  This is elaborated on in the next figure.
}
\label{fig: atom_moment}
\end{figure}

Note that our use of the asymptotic expressions for the different many-body functions aided us in reducing the effort of computation only for problems that can be solved along the imaginary axis.  This includes the Falicov-Kimball model (for static properties) and Hirsch-Fye quantum Monte Carlo techniques for other models (like the Hubbard model).  There does not appear to be any simple use of these sum rules within real-frequency-based approaches like the numerical renormalization group.
Of course, all of these results can also be applied to the homogeneous case.

\begin{figure}[ht]
\centerline{\includegraphics [width=3.2in, angle=0, clip=on]  {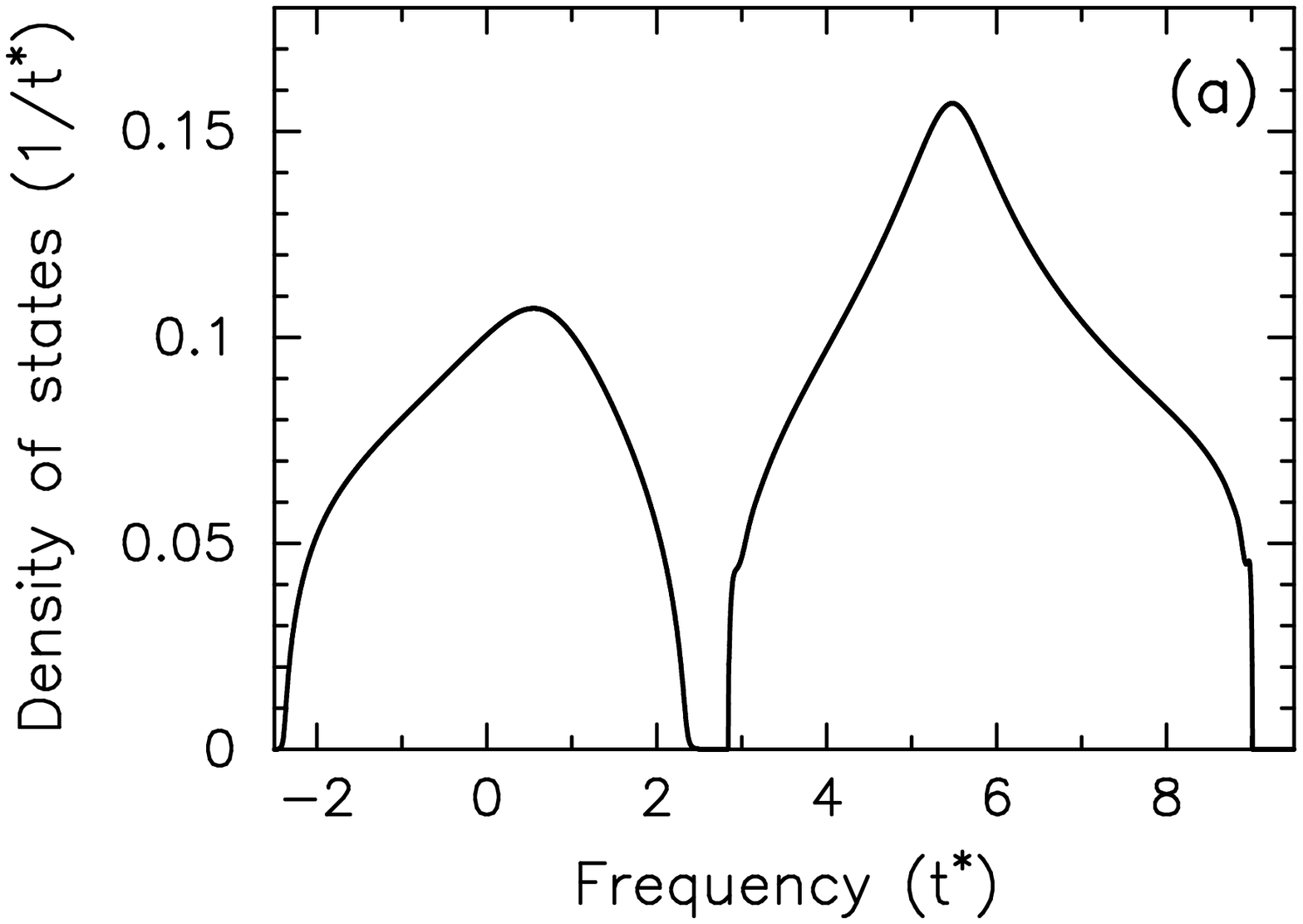}
\includegraphics [width=3.04in, angle=0, clip=on]  {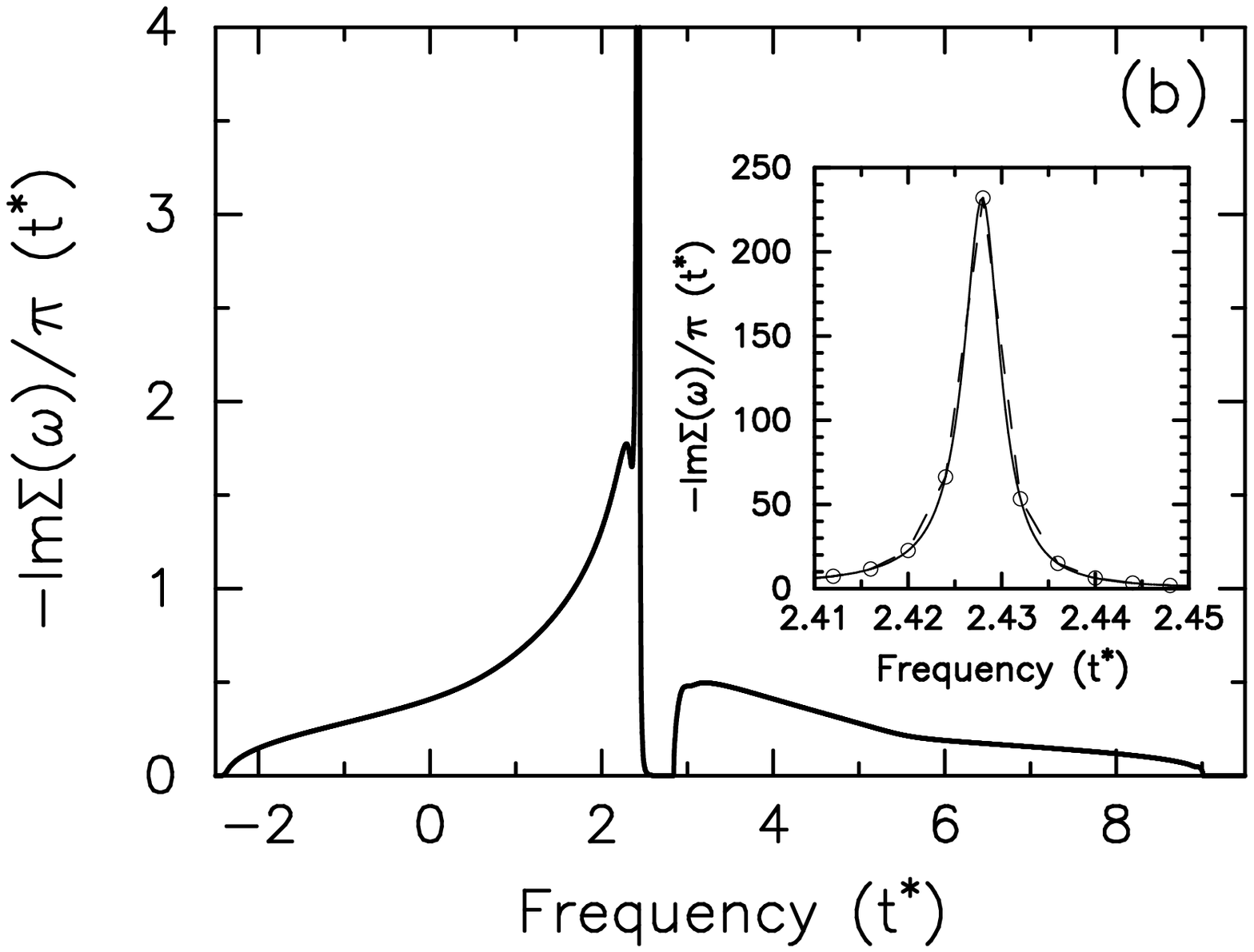}
}
\caption[]{
Density of states (a) and $-{\rm Im}\Sigma_i(\omega)/\pi$ (b) for the central site of the lattice with the same
parameters as the preceeding figure.  Note how the DOS is smooth and rather featureless, so the sum rules work to high accuracy, while the self-energy has sharp features, which lead to less accurate moments. In the inset of panel (b), we show the large peak in the self-energy; the data with the solid line has a small step size of $0.0003t$, while the dashed lines and the circles correspond to a step size of $0.004t$. Note how the smaller step size is smoother.}
\label{fig: dos_se}
\end{figure}

In Fig.~\ref{fig: atom_moment}, we plot false grayscale images of two moment sum rules for each lattice site of a $51\times 51$ square lattice.  This is a system close to phase separation with parameters $U=5$, $T=0.2$, and a harmonic trap with a characteristic length scale of 30 lattice spaces for both the light and the heavy particles. 
Note how the errors, on the whole, are rather small.  In the imaginary-axis calculation, which is used to determine the chemical potentials so that we have approximately 625 light and 625 heavy atoms on the lattice, we employ the use of the moment sum rules to sum the tails of the Matsubara frequency calculations, which reduces the computational time by about a factor of 10 for the same accuracy. In this real axis calculation, we use a frequency grid with a step size of $0.004t$ that runs from $-9.6t$ to $9.6t$.  This step size does not allow us to pick up fine structure smaller
than the step size.  It also cannot pick up spectral weight lying outside of the bounds.  Both of these issues can
cause inaccuracies in calculations, especially when a system begins to order or phase separate; they don't enter too significantly for this example though; when we repeat the calculations with a smaller step size of $0.0003t$, we find the error in the zeroth self-energy moment is reduced from 1.5\%  to 0.025\% for the central site of the lattice.  The temperature we use here
is high enough that there is no order, and the calculations are under good control.  In any case, we show the spectral function and $-{\rm Im}\Sigma_i(\omega)/\pi$ for the central site of the lattice in Fig.~\ref{fig: dos_se}, where the relative error for the zeroth moment of the self-energy is about 1.5\%, and the relative error for the Green function is about 0.01\%.  Notice the sharp peak in the self-energy which leads to the higher error, while the Green function is quite smooth, and hence has very small errors. As the step size is reduced, the errors in the moments are also reduced,
indicating that these errors are arising predominantly from the discretization size of the real frequency axis and
the structure of the sharp features in the functions.

\begin{figure}[t]
\centerline{\includegraphics [width=2.9in, angle=0, clip=on]  {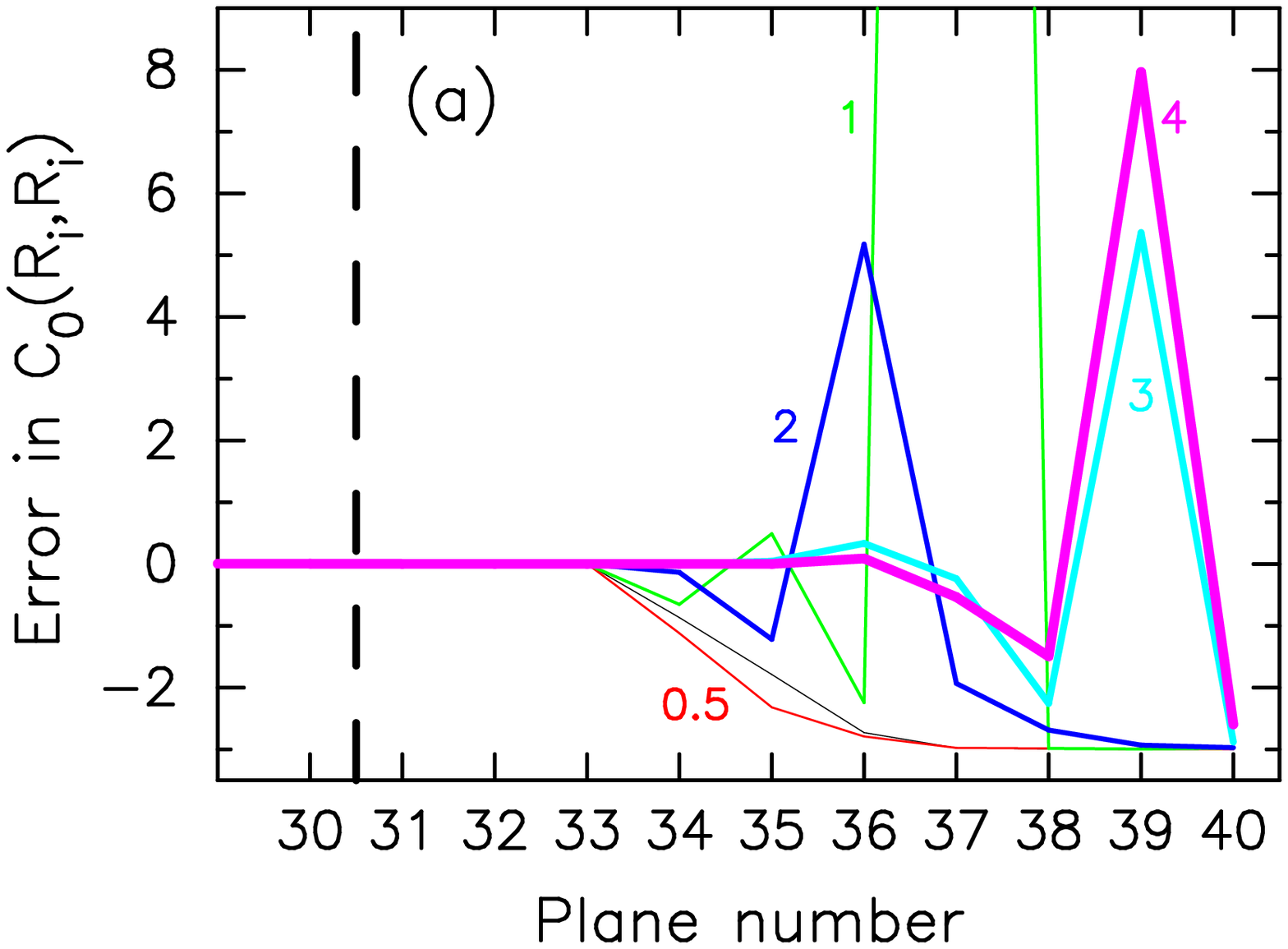}
\includegraphics [width=3.2in, angle=0, clip=on]  {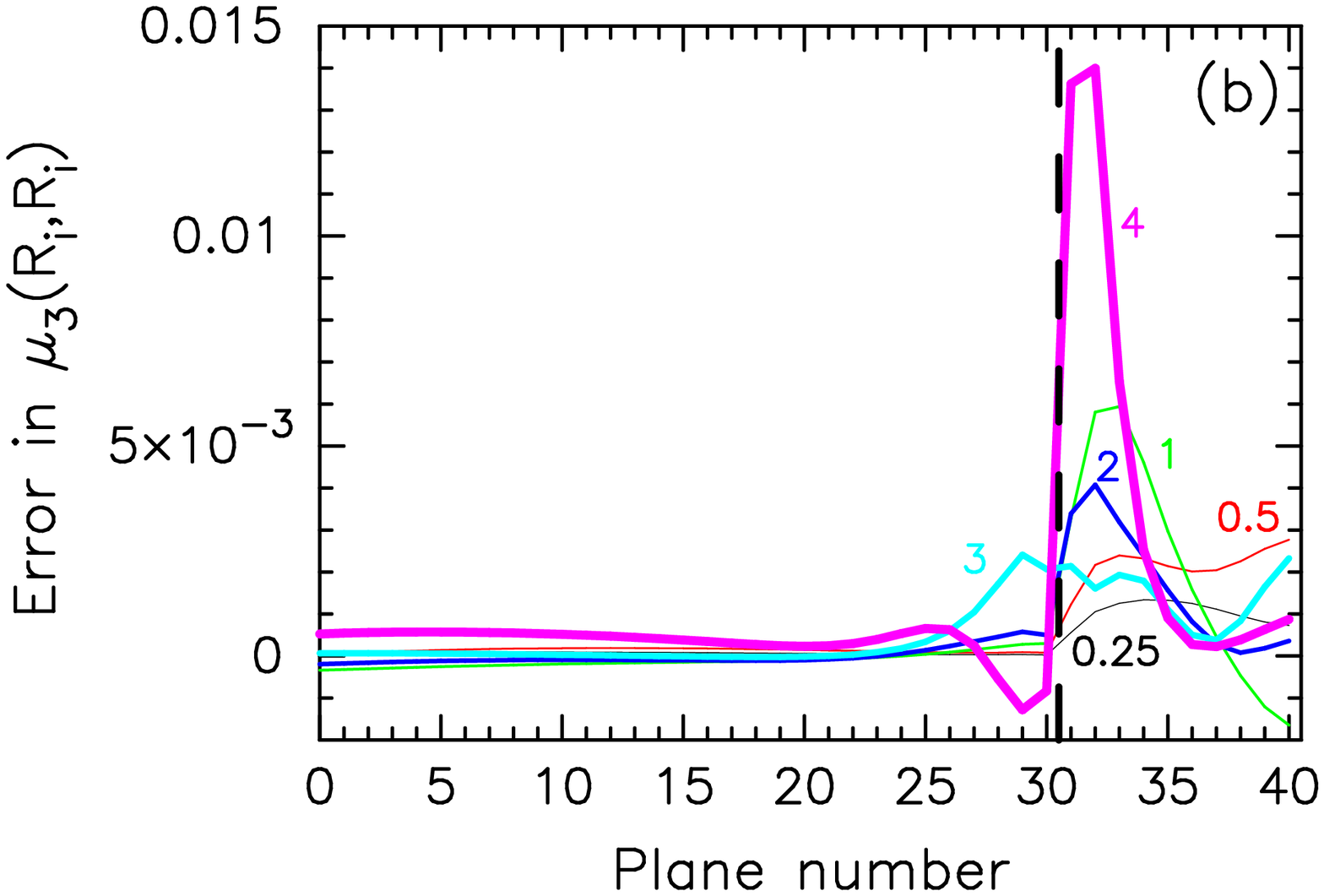}}
\caption[]{
(Color online) plot of the absolute error of the (a) zeroth self-energy
moment and the (b) third spectral function moment for a multilayered inhomogeneous system described by the 
Falicov-Kimball model with electronic charge reconstruction. There are 30 self-consistent ballistic metal planes to the right and the left of the 20 plane thick barrier which has $U=6$ and half-filling for the heavy electrons. The temperature is $1/\beta=0.1$, and the shift in the band centers $\Delta E_f$ labels the different figures. Note how the errors for the self-energy are larger, while for the Green function the errors are small.  This is because our grid size is too coarse to properly pick up the weight of the narrow peak in the self-energy.  The relative error for the third moment of the DOS is less than 0.1\% in the barrier; in the metallic leads, the third moment gets very small, and the absolute error arises primarily from the discretization of the numerical quadrature. The dashed line indicates the interface between the metal and the insulating barrier.
}
\label{fig: multi_moment}
\end{figure}

We also examine the sum rules for inhomogeneous multilayered systems.  In cases where there is no electronic charge
reconstruction, we have found that our data satisfies all of the sum rules to high accuracy (typically better than
0.01\%) except for cases with an insulating barrier where the self-energy develops a sharp peak
at low frequency.  Our frequency grid in previous calculations was sometimes too large to properly extract the 
zeroth moment sum rule for the self-energy, and errors could become very large because the numerical quadrature
is greatly overestimating the weight within the sharp peak near $\omega=0$.  In cases where the peak is not so sharp,
we once again find excellent agreement.  A more challenging case, though, is a case where there is an electronic
charge reconstruction, because the calculations become much more difficult numerically in this case, and we
usually need to introduce a finite broadening into the calculation to be able to estimate the local DOS on each plane.  Hence, it is much more interesting to examine these cases for calculations of the sum rules.

\begin{figure}[ht]
\centerline{\includegraphics [width=3.2in, angle=0, clip=on]  {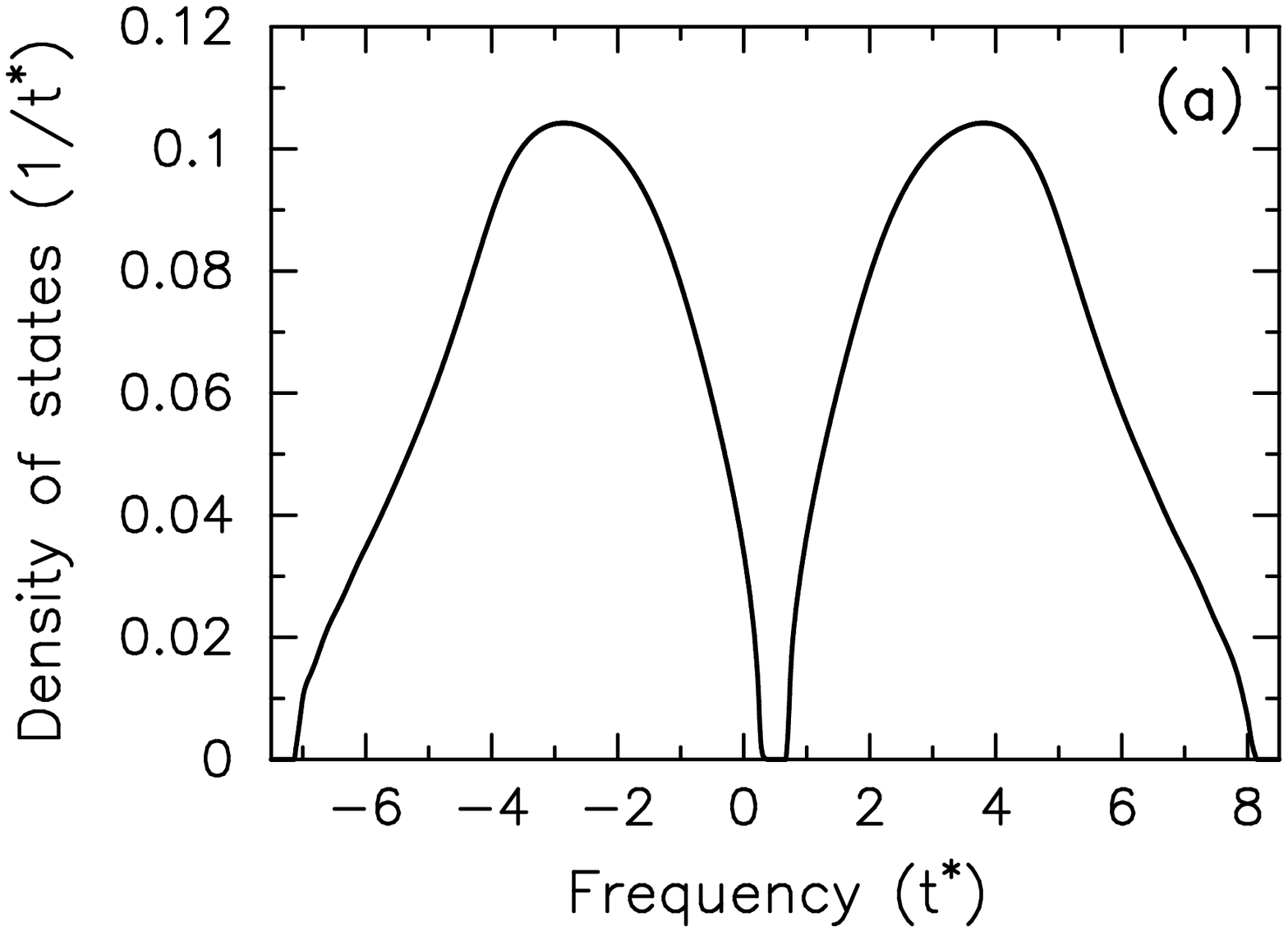}
\includegraphics [width=3.04in, angle=0, clip=on]  {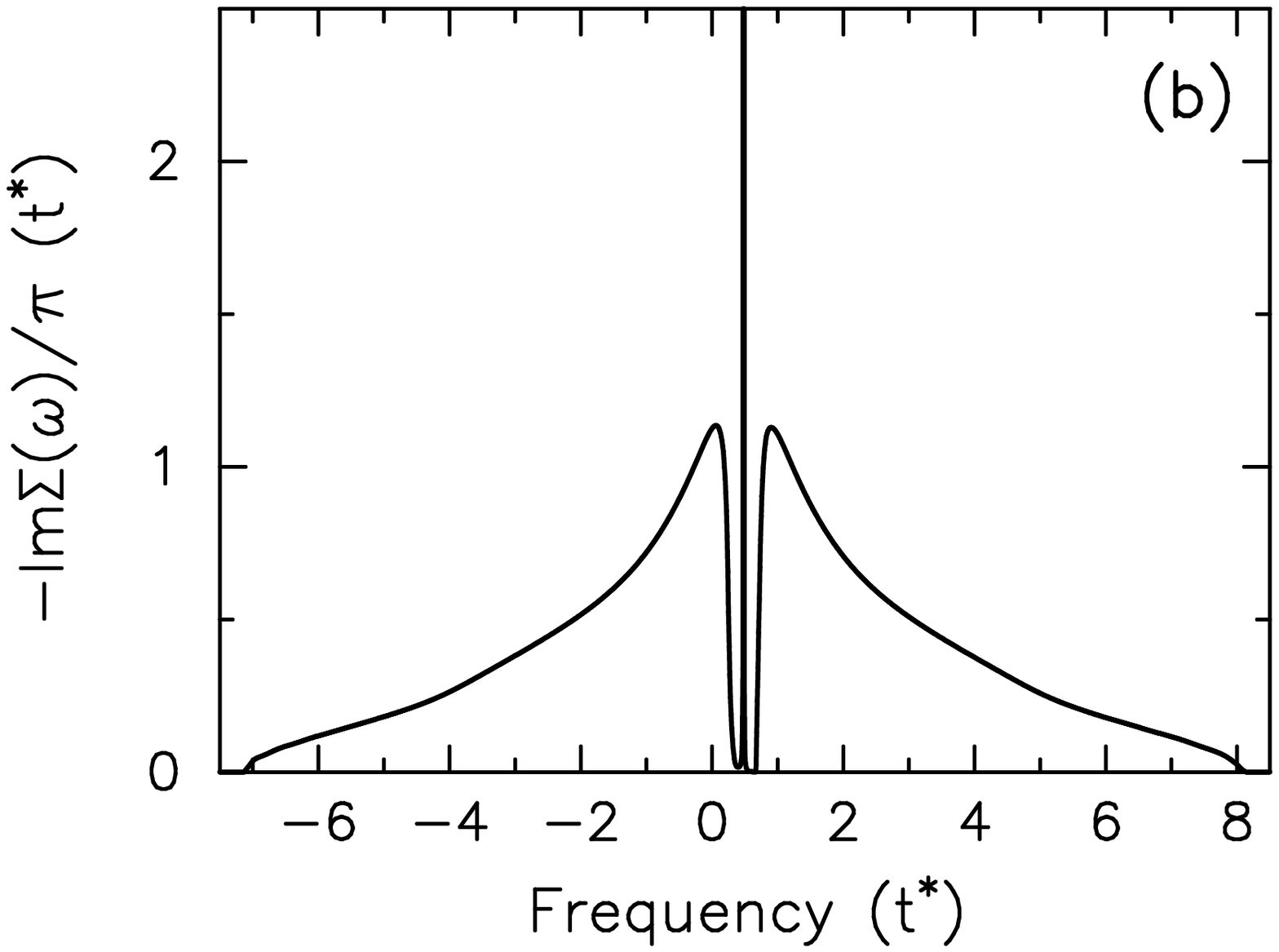}
}
\caption[]{
Density of states (a) and $-{\rm Im}\Sigma_i(\omega)/\pi$ (b) for the plane with the largest error (plane number 37 with $\Delta E_f=1$) in the preceeding figure.  Note how the DOS is once again smooth and rather featureless, so the sum rules work to high accuracy, while the self-energy has a sharp peak (maximum amplitude is 8000), which leads to less accurate moments.}
\label{fig: dos_se_multi}
\end{figure}

In Fig.~\ref{fig: multi_moment}, we plot the absolute errors of a self-energy moment and a spectral moment (the other self-energy moment appears similar, while the other spectral moments had much smaller errors). The system consists of a semi-infinite bulk ballistic metal attached to a sandwich of 30 ballistic metal planes, 20 Falicov-Kimball model planes and 30 ballistic metal planes, so the calculations are always for a thermodynamic limit system. Both the metallic leads and the barrier are at half filling, with a common chemical potential.  We shift the center of the band of the barrier by the amount $\Delta E_f$ and solve for the electronic charge reconstruction with a screening length of a few lattice spacings and a temperature of $1/\beta=0.1$. The imaginary axis solver did not use the summation of the tails, as we are using old data from Ref.~\onlinecite{freericks_multilayer_thermal}.  The real-axis solver worked with a grid step size of $0.01t$ and ranged $\omega$ from $-11t$ to $11t$.  One can see that, similar to the cold atom example above, here we also see errors which are much larger for the self-energy than for the spectral function (we only show the barrier planes for the self-energy moment, since the self-energy vanishes in the metal). This arises for the same reason as before, but is more acute here, since the step size in frequency is larger.  This is shown in Fig.~\ref{fig: dos_se_multi} for the plane with the largest error in the self-energy. Note how once again the DOS is smooth, which is why the moments are so accurate, but the self-energy has a narrow peak, whose weight is overestimated with the coarse grid used in the calculation.

\begin{figure}[ht]
\centerline{\includegraphics [width=3.5in, angle=0, clip=on]  {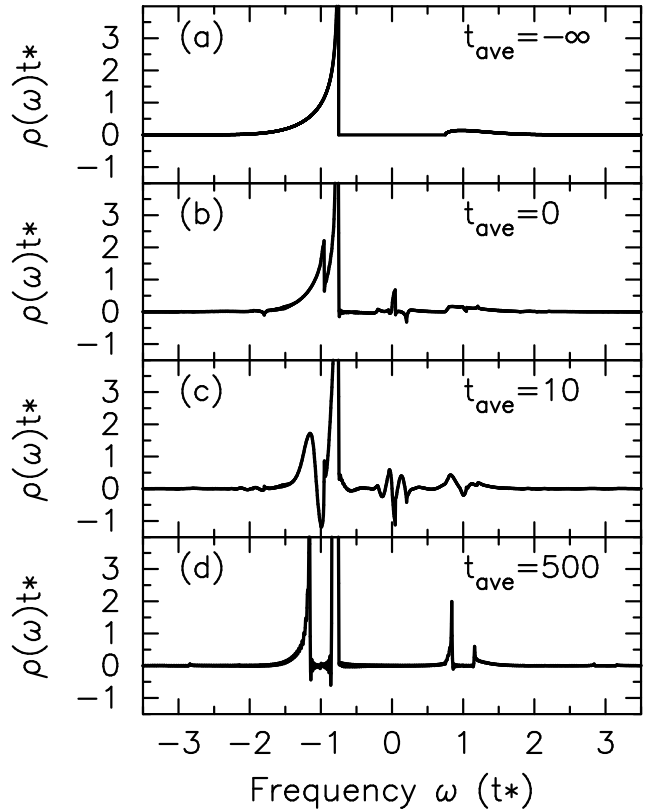}
}
\caption[]{
Density of states for the $A$ sublattice at $U=1.5$ with an electric field of strength $E=1$ turned on at time
$t=0$. The different panels correspond to different average times.  Note how the main structure of the DOS in equilibrium, which
consists of the singular peak and the finite peak is modified at the {\it odd} Bloch frequencies here to create
additional structures that look reminiscent of the DOS of an ordered system (one can see small peaks near $\omega=\pm 3$ too).  Modifications at the {\it even}
Bloch frequencies can only be seen at short times.}
\label{fig: dos_cdw_u=1.5}
\end{figure}

Spectral moment sum rules for ordered phases, like a charge-density-wave phase, have already been performed in
equilibrium\cite{optical_conductivity} for the Falicov-Kimball model, and agree with the exact results to high accuracy.
The spectral moment sum rules for the nonequilibrium case in a homogeneous system with a uniform electric field have
also been verified to high accuracy\cite{turkowski_freericks1,turkowski_freericks2}. In addition, in the rapid quench work of Ref.~\onlinecite{eckstein}, the retarded Green function is given by the equilibrium Green function of the particular value of the interaction for each average time.  Hence, the moments, which hold in equilibrium, continue to hold in nonequilibrium with an interaction quench. Here we examine a nonequilibrium case at half filling for both localized and itinerant electrons with charge density wave order in the presence of a uniform electric field at zero temperature.  This system can also be solved exactly within dynamical mean-field theory, because the self-energy vanishes or is equal to $U$ and it has no damping ({\it i. e.}, it is real when expressed in the frequency-average time representation).  Details of that work will appear elsewhere\cite{shen_freericks} and follow closely the derivation of the nonequilibrium Green function for noninteracting electrons on a lattice\cite{turkowski_freericks_bloch}, but with some added complications due to the need for time-ordered products because of the CDW order. The local retarded Green function at half-filling with $\mu=U/2$ satisfies\cite{shen_freericks}
\begin{equation}
 G^R_{ii}(t_1,t_2)=-i\theta(t_1-t_2)\sum_{\bf k}\exp\left [ \mathcal{U}_{11}({\bf k},t_1,t_2)+\mathcal{U}_{22}({\bf k},t_1,t_2)
\pm \mathcal{U}_{12}({\bf k},t_1,t_2)\pm \mathcal{U}_{21}({\bf k},t_1,t_2)\right ],
\end{equation}
where the sum over momentum is over the CDW Brillouin zone, which satisfies $\epsilon_{\bf k}\le 0$, and the plus sign is for $i\in A$ sublattice and the minus sign for $i\in B$ sublattice.  The time evolution operator $\mathcal{U}$ is a time-ordered product
\begin{equation}
\mathcal{U}({\bf k},t_1,t_2)= \mathcal{T}_t\exp\left [ i\int_{t_2}^{t_1} dt \left ( \begin{array}{c c} -|\epsilon_{{\bf k}-e{\bf A}(t)}|&\frac{U}{2}\\\frac{U}{2}&|\epsilon_{{\bf k}-e{\bf A}(t)}|\end{array}\right )\right ].
\end{equation}
We used the Peierls' substituted band structure with $\epsilon_{\bf k}=\lim_{d\to\infty}t^*\sum_{i=1}^d\cos ({\bf k}_ia)/\sqrt{d}$ the band structure and ${\bf A}(t)$ the vector potential, which is turned on at time $t=0$ [${\bf A}(t)=-\theta(t){\bf E}t$]. The electric field is chosen to lie along the diagonal direction. The time-ordered product can be calculated directly for numerical work by employing the Trotter formula on the corresponding $2\times 2$ matrices.

\begin{figure}[ht]
\centerline{\includegraphics [width=3.5in, angle=0, clip=on]  {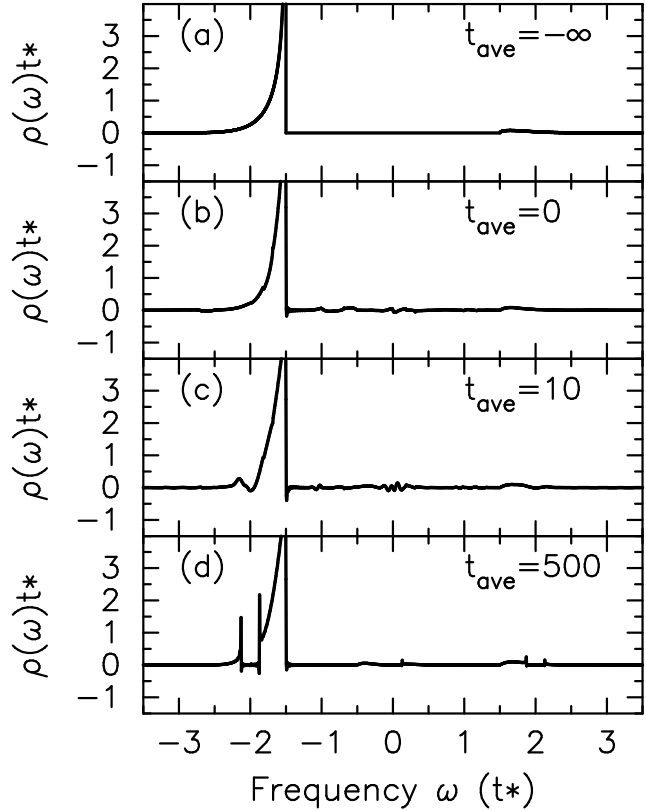}
}
\caption[]{
Density of states for the $A$ sublattice at $U=3$ with an electric field of strength $E=1$ turned on at time
$t=0$. The different panels correspond to different average times.  Note how the main structure of the DOS in equilibrium, which
consists of the singular peak and the finite peak is modified at the {\it even} Bloch frequencies here to create
additional structures that look reminiscent of the DOS of an ordered system (small peaks are near $\omega=0$ too). Modifications at the {\it odd}
Bloch frequencies can only be seen at short times.}
\label{fig: dos_cdw_u=3}
\end{figure}

Now we report on the nonequilibrium sum rules for the CDW phase.  In this case the moments of the local Green functions on each sublattice satisfy the following: (i) the zeroth moment is 1; (2) the first moment is $\pm U/2$; (iii) the second moment is $1/2+U^2/4$; and (iv) the third moment is $\pm U/4\pm U^3/8$ (the plus or minus signs correspond to the $A$ or $B$ sublattice).  While, in theory, one can numerically calculate these moments by first finding the Green functions as functions of time, converting to average and relative time, Fourier transforming the relative time to a frequency, and finally calculating the moment sum rule by integrating over frequency (see below), this approach runs into a number of serious challenges.  The most critical one is that the equilibrium DOS, when there is no field present, has a divergence in it that goes like an inverse square root of frequency.  Such a singularity requires an infinite time domain to properly find the Fourier transform (because the function has an amplitude that decays like a power law in time), but our numerical calculations are always truncated to a finite range, so the Fourier transform has the singularity smoothed over and the truncation can lead to unphysical oscillations in the DOS as a function of frequency (due to the presence of a sharp cutoff in time).  Furthermore, we calculate on a discrete grid in time, which can have further effects on the Fourier transform, especially for high enough frequencies. Finally, the results, particularly at large times, are sensitive to the number of energy points in the integration grid over the two energies. All of these challenges make it much more useful to directly calculate the results for the moments by evaluating the derivatives of the Green functions  as functions of time.  This can actually be done analytically for the form of the Green function (because of the time-ordered products), and it produces exactly the requisite moments.  In addition, we can do it numerically in the time domain, and we verify that the sum rules are satisfied to an accuracy much smaller than the step size in time  (when evaluated in the time domain via numerical differentiation).  

Even though the numerical calculation of the DOS in the CDW phase is challenging, as explained above, we present results for this calculation in Figs. \ref{fig: dos_cdw_u=1.5} and \ref{fig: dos_cdw_u=3} for the $A$ sublattice DOS. The parameters chosen are an electric field equal to 1, and turned on at time $t=0$, and interaction strength $U=1.5$ and $U=3$ (we have a time domain cutoff running from $-500$ to $500$ with a time step of $0.0025$). We compare a number of different average time results including (a) the equilibrium result $t_{\rm ave.}\rightarrow -\infty$, (b) $t_{\rm ave.}=0$, (c) $t_{\rm ave.}=10$, and (d) $t_{\rm ave.}=500$. We do not use the ``equilibrium'' results from the real time calculation because our finite time-domain cutoff leads to spurious oscillations. In addition, the DOS is already modified at $t_{\rm ave.}=0$ because the relative time Fourier transform involves a large number of points in time where the field is on, and because the Green function has structure with such long-ranged tails in time, one can see an effect even before the average time where the field has been turned on. When the field is turned on, the DOS naturally changes shape (and the square-root singularity appears to be smoothed into a finite peak), but the changes are much smaller than in the normal phase.   Surprisingly, we predominantly see structure at either odd Bloch frequencies or even Bloch frequencies, but not both, and the structure does not look like a broadened, and split delta function as in the normal phase, but instead looks more like some kind of ordered phase gap structure, but the features
can be quite small in some cases.  It is clear we have not yet fully reached the steady state, but it also appears clear what the steady state DOS will eventually look like. Finally, we comment that if we take the DOS as functions of frequency (our frequency range is chosen to run from $-10<\omega<10$), multiply by the appropriate power of frequency and integrate to check the sum rules, then all sum rules except the second moment are satisfied to better than 1\%.  The second moment is worse, because we get significant contributions from the noisy tails of the DOS which don't cancel as they do for the odd moments.  If we put the frequency cutoff closer to $|\omega|<5$, then the second moment accuracy increases dramatically.  But we know that this is the least accurate way to check the moment sum rules.

\section{Conclusions}

In this work, we have shown how to determine exact spectral moment sum rules for the retarded Green functions and self-energies of inhomogeneous strongly correlated systems.  While our analysis is quite general, for concreteness, we provided explicit results for a combined Hubbard-Falicov-Kimball model.  We envision these results can be applied to strongly correlated multilayers, ultracold atomic systems in traps, or strongly correlated materials with disorder (although we did not discuss the last case in much detail). The sum rules can be used to gain qualitative information about the many-body solutions, benchmark numerical calculations, be used to evaluate the tails of infinite sums (or infinite products) allowing for smaller energy cutoffs, or be used to improve the accuracy of Hirsch-Fye quantum Monte Carlo approaches by determining the short-imaginary-time behavior exactly.  We provided a full derivation of the sum rules in equilibrium, and then discussed a number of different nonequilibrium situations appropriate for multilayers, for cold atom systems, and for ordered phase systems (such as a CDW).  In the case of multilayers, the change in the local electrical potential energy, induced by an electronic charge reconstruction, modifies the sum rules, but the additional scalar potential, that creates an electric field to drive current through the device, does not.  This motivates one to decouple the description, using a scalar potential and Poisson's equation in a semiclassical analysis to determine the modified Hamiltonian due to the electronic charge reconstruction, but use a time-dependent vector potential to describe the electric field that drives current through the system. In the case of cold atoms, we discussed nonequilibrium situations corresponding to pulling the lattice in the presence of a static trap, rapidly changing the location of the trap and examining the transient response, and changing the interaction strength (say due to a Feshbach resonance, by changing the magnetic field) and examining the response to an interaction quench. For the ordered phase, we examined the CDW state of the Falicov-Kimball model at zero temperature, with an uniform electric field turned on at a specific time. We feel these sum rules will have a broad application across a number of different systems and calculations.

It is obvious that a similar exercise can be carried out for bosonic systems.  These will have most relevance for
cold atom problems, where bosonic atoms are readily available within the alkali family.  The sum rules for bosons
become more complicated, especially at higher orders, because we used identities such as $c_i^\dagger c^{}_i c^\dagger_ic^{}_i=c_i^\dagger c^{}_i$ which hold only for fermionic systems.  We are currently working on the
generalization of these sum rules to bosonic problems which will be presented elsewhere. Of particular interest is a Falicov-Kimball model with light fermions and heavy bosons, which is realized in K-Rb mixtures used to make dipolar molecules if the system is placed on a deep enough optical lattice.

\acknowledgments

J.~K.~F. acknowledges support from the National Science Foundation under grant number DMR-0705266.
Supercomputer time was provided by the ERDC, ASC, and ARSC supercomputer centers of the HPCMP of the DOD. The largest real-axis cold atom calculations were performed during a capabilities application project in 2008 on the Cray XT5 at ARSC.
V.T. acknowledges support by the Department of Energy under grant number DE-FG02-07ER15842. We also acknowledge useful conversations with N. Bl\"umer, M. Kollar, H. Krishnamurthy and W. Shen.

\end{document}